\documentclass{article}
\usepackage{spconf,amsmath,graphicx,amssymb,multirow,booktabs,lineno,pgfplots}

\title{Speaker Change Detection for Transformer Transducer ASR}
\name{Jian Wu, Zhuo Chen, Min Hu, Xiong Xiao, Jinyu Li}
\address{Microsoft, One Microsoft Way, Redmond, WA, USA}
\begin{document}
\ninept
\maketitle
\begin{abstract}

Speaker change detection (SCD) is an important feature that improves
the readability of the recognized words from an automatic speech recognition (ASR) system by breaking the word sequence into paragraphs at speaker change points.
Existing SCD solutions either require
additional ensemble for the time based decisions and recognized
word sequences, or implement a tight integration between ASR and
SCD, limiting the potential optimum performance for both tasks.
To address these issues, we propose a novel framework for the SCD task, where an additional SCD module is built on top of an existing Transformer Transducer ASR (TT-ASR) network. 
Two variants of the SCD network are explored in this framework that naturally estimate speaker change probability for each word, while allowing the ASR and SCD to have independent optimization scheme for the best performance.
Experiments show that our methods can significantly improve the F1 score on LibriCSS and Microsoft call center data sets without ASR degradation, compared with a joint SCD and ASR baseline.

\end{abstract}
\begin{keywords}
Speaker Change Detection, Transformer Transducer, E2E ASR, F1 Score
\end{keywords}
\vspace{-0.7em}
\section{Introduction}
\vspace{-0.5em}


Speaker change detection (SCD) is a task to estimate the speaker transition point in an audio stream that contains potentially more than one active speaker. 
In modern speech processing applications, a speaker change detection module usually provides two important features. 
Firstly, accurate speaker change signal is crucial for improving the display format of the conversational recognition system, where the displayed transcription is expected to start a new line as the active speaker changes.
Secondly, SCD serves as an important front end processing for speaker diarization applications \cite{anguera2012speaker,park2022review}, where the input speech is first segmented by speaker change points before further processing, such as clustering.


Existing SCD systems can be roughly categorized into two types, the time based detection \cite{bredin2017speaker,hruz2017convolutional,ge2017speaker,sari2019pre} and word based detection \cite{xia2022turn,fan2022token}. 
In the time based detection, SCD is usually designed to rely on pure acoustic features from input speech, and the decision is estimated on a frame basis. Although they enjoy a higher decision resolution, the time based decision usually suffers from three limitations. 
Firstly, it's hard to locate the accurate speaker boundaries due to the impact of the silence or noises \cite{hruz2017convolutional}. 
Secondly, when combining with ASR systems, 
a post processing step is required to insert the speaker change label into the recognized word sequence as the detected time based change points may not fall between word boundaries.
Finally, only acoustic information is explored although the semantic information has been shown to be beneficial for the SCD and diarization task \cite{park2018multimodal,india2017lstm,flemotomos2019linguistically}.

Unlike the time based detection, the word based SCD directly estimates the speaker change point between words. 
In this way, the SCD decision naturally aligns with the ASR word sequences, removing the synchronization steps required in the time based system. 
More importantly, this method allows SCD to access both acoustic and semantic information from the ASR module, leading to higher potentially performance.
On the other hand, with the recent advances of the E2E ASR techniques \cite{li2020comparison,he2019streaming,li2022recent}, various tasks including SCD \cite{xia2022turn} have been integrated with the ASR in the past, such as endpoint detector \cite{li2020towards}, turn taking \cite{chang2022turn},  speaker diarization \cite{shafey2019joint,mao2020speech} and multi-talker recognition \cite{kanda2022streaming}, and shown the competitive results compared with the modular approaches.



Although the word based SCD systems have shown promising results, existing joint SCD-ASR \cite{xia2022turn} solution often suffers from two limitations. 
Firstly, it's hard to reach the optimal performance for both tasks in a joint model architecture. 
In our investigation, a higher word error rate (WER) is often observed when additional tuning to improve SCD is included during training.
Secondly, as SCD network detects the acoustic change in speaker characteristics, longer latency is needed to ensure the robustness of SCD module. 
However, due to the streaming request of the ASR applications, the joint SCD-ASR model often adopts a low latency setting, limiting the performance for the SCD module.

To address these limitations, in this work, we propose a speaker change detection method based on a existing Transformer Transducer ASR (TT-ASR) \cite{chen2021developing} network.
Instead of directly incorporating SCD into ASR decoding, our model
introduces an additional network for SCD on top of the TT-ASR.
Given the partial ASR decoding word sequence, the SCD network predicts the word-level speaker change probabilities. 
Specifically, two implementations of proposed framework are investigated, namely \textbf{SCD-ENC} and \textbf{SCD-DEC}, depending on whether the explicit alignments of recognized word sequence from TT-ASR are available. 
Based on the proposed SCD model, a chunk-wise streaming processing pipeline with flexible latency during inference is introduced. 
We evaluate the models in two data sets and the results show that compared to a joint SCD-ASR baseline, the proposed SCD models achieve significant better F1 scores in both sets without losing ASR accuracy.  

The rest of the paper is organized as follows. The TT-ASR and SCD methods are introduced in Section 2 and 3, respectively. Experimental setups including the data, model, and training details are described in Section 4. The results are discussed in Section 5. Section 6 mentions the future work and concludes the paper.

\vspace{-0.7em}
\section{Transformer Transducer ASR}
\vspace{-0.5em}

\begin{figure*}[!tbp]
\centering
\vspace{-4mm}
\includegraphics[width=0.9 \textwidth]{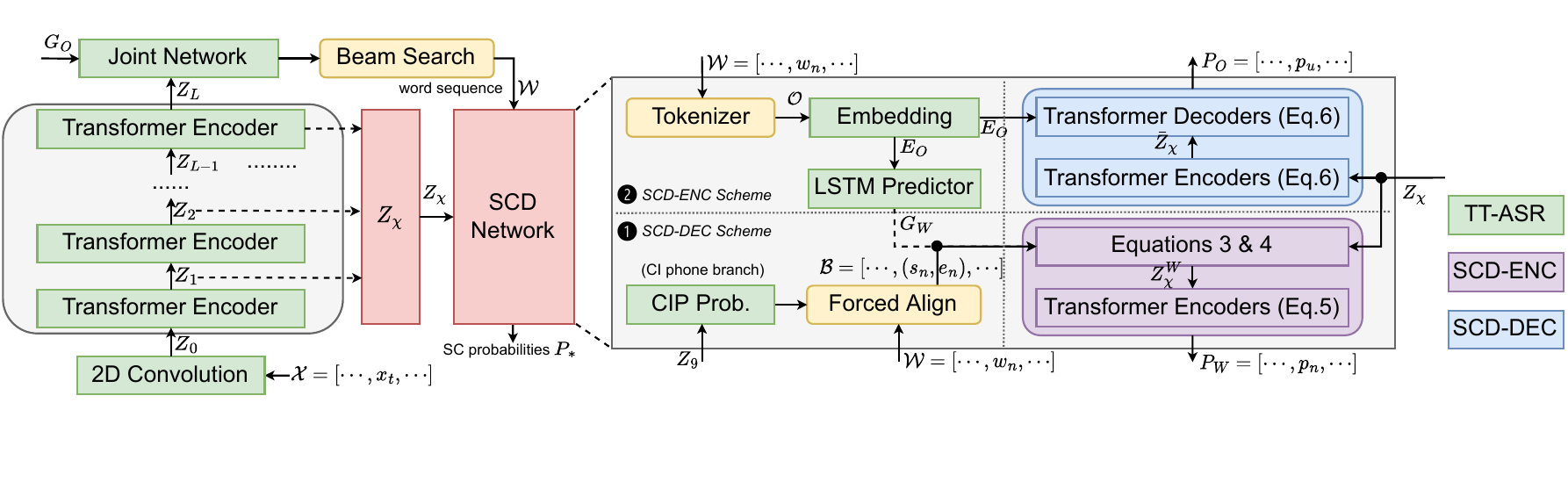}
\vspace{-8mm}
\caption{Illustration of our TT-ASR based SCD method. The green blocks indicate the pure ASR modules and the yellow round squares represent runtime algorithms. SCD-ENC and SCD-DEC implementations are shown by blue and purple blocks in the right, respectively.}
\vspace{-4mm}
\end{figure*}


Our TT-ASR is based on the neural transducer structure \cite{graves2012sequence} while a stack of the Transformer \cite{vaswani2017attention} encoder is used with a streaming solution \cite{chen2021developing} for high-accuracy and low-latency. 
The output of the $i$-th ($1 \leqslant i \leqslant L$) encoder layer $Z_i$ is calculated with the pre-normalization as in the following equations
\vspace{-0.6em}
\begin{equation}
\begin{aligned}
    \bar{Z}_{i-1} & = {Z}_{i-1} + \mathtt{MHSA}_i \left(\mathtt{ln}({Z}_{i-1}) \right), \\
    Z_i & = \bar{Z}_{i-1} + \mathtt{FFN}_i \left(\mathtt{ln}(\bar{Z}_{i-1}) \right),
\end{aligned}
\vspace{-0.4em}
\end{equation}
where $\mathtt{ln}(\cdot)$ denotes the layer normalization. $\mathtt{MHSA}_i(\cdot)$ and $\mathtt{FFN}_i(\cdot)$ represent the $i$-th layer multi-head self-attention (MHSA) block and feed-forward network (FFN), respectively. The learnt relative position encoding is adopted in the MHSA implementation. $Z_0$ is calculated as $Z_0 = \mathtt{CONV}(\mathcal{X})$ given the $T$-frame acoustic feature $\mathcal{X} = [x_0, \cdots, x_{T - 1}]$ and $\mathtt{CONV}(\cdot)$ represents a 2D convolutional subsampling layer which is used to reduce the frame rate before the Transformer encoder. 

Regarding to the decoder in TT-ASR, we use long short-term memory (LSTM) as the prediction network and a linear layer as the joint network. Given the estimated non-blank token $o_u$ and hidden state $h_u$, the predictor's output $g_{u+1}$ at $u+1$ step is wrote as
\vspace{-0.6em}
\begin{equation}
g_{u+1}, h_{u+1} = \mathtt{LSTM} \left(\mathtt{Embed}(o_u), h_u \right)
\vspace{-0.6em}
\end{equation}
and $\mathtt{Embed}(\cdot)$ means the token embedding layer. The one-best decoding token (sentence piece) sequence $\mathcal{O} = [o_0, \cdots, o_{U - 1}]$ (or $N$-word sequence $\mathcal{W} = [w_0, ..., w_{N - 1}]$) of length $U$ are obtained by the beam search algorithm over the probability distribution output from the joint network.

\vspace{-0.7em}
\section{Method}
\vspace{-0.5em}

\subsection{Feature Design}
\vspace{-0.3em}


The hidden representations extracted from ASR encoder are used as the basic acoustic feature for SCD network.
Apart from $Z_{i}$, the weighted-sum $Z_{\omega} = \sum_{i=1}^L \omega_{i-1} Z_i$ over the encoder layer outputs $\{ Z_{1}, \cdots, Z_{L} \}$ is also investigated. Similar to \cite{chen2022wavlm}, the learnable weights $\omega$ are constrained as $\sum_i \omega_i = 1$. Regarding to the semantic features, different representations from TT-ASR are adopted in different implementations, which will be discussed in the next section.


\vspace{-0.5em}
\subsection{SCD Network Implementation}
\vspace{-0.3em}

Taking acoustic feature $Z_\chi \in \{Z_1, \cdots, Z_L, Z_\omega\}$ and word sequences $\mathcal{W}$ as input, the SCD decision network is designed to align the information from both modality, explicitly or implicitly, and estimates the final speaker change point for each word, as shown in Figure 1. 
Specifically, we explore two implementations, SCD-ENC and SCD-DEC, depends on whether alignment based word boundaries \cite{zhao2021addressing} are available through our TT-ASR model. 
In SCD-ENC, an explicit forced alignment step is included to obtain the word boundary from recognized word sequence. The acoustic features $Z_\chi$ are then averaged and transformed into the word based representations $Z_\chi^W$, which is further processed by a stack of transformer encoders to produce the final speaker change decision.
On the other hand, in SCD-DEC, we utilize Transformer decoders, which include the source-target attention mechanism between the token sequences $\mathcal{O}$ and acoustic features $Z_\chi$, to directly estimate the word based speaker change without explicit word boundary information.

\textbf{SCD-ENC} As we have the ASR decoding word sequence $\mathcal{W} = [w_0, ..., w_{N - 1}]$ as well as their word boundaries $\mathcal{B} = [(s_0, e_0), \cdots, (s_{N-1}, e_{N-1})]$ through forced alignment, a stack of Transformer encoder layers are used to predict the speaker change probability of each word, as shown in the purple blocks of Figure 1. The audio embedding $z_{\chi}^{w_n}$ of word $w_n$ is calculated by averaging the corresponding parts in $Z_{\chi}$ matrices given the word boundary $(s_n, e_n)$:
\vspace{-0.5em}
\begin{equation}
z_{\chi}^{w_n} = \mathtt{Mean} \left(Z_{\chi}^{[s'_n, e'_n]} \right),
\vspace{-0.5em}
\end{equation}
where $s'_n$ and $e'_n$ is calculated based on the subsampling rate and kernel size of the convolution layer.
We can further utilize the TT-ASR's predictor outputs $G_O = [g_0, \cdots, g_{U - 1}]$ to compensate the missing semantic information in $z_{\chi}^{w_n}$. In this work, we use the last token representation of the word as the text feature of each word and concatenate it with the original $z_{\chi}^{w_n}$ to form a new one:
\vspace{-0.5em}
\begin{equation}
z_{\chi}^{w_n} = \mathtt{Concat} \left( \left[z_{\chi}^{w_n}, g_{w_n} \right] \right),
\vspace{-0.5em}
\end{equation}
where $g_{w_n}$ is the predictor's last token representation for word $w_n$. Finally the audio embedding sequences $Z_{\chi}^W = [z_{\chi}^{w_0}, \cdots, z_{\chi}^{w_{N-1}}]$ are feed to Transformer encoder layers to estimate the word-level speaker change probability $P_W = [p_0, \cdots, p_{N - 1}]$:
\vspace{-0.5em}
\begin{equation}
P_W = \sigma \left(\mathtt{TransformerEnc} (Z_{\chi}^W) \right),
\vspace{-0.5em}
\end{equation}
where $\sigma(\cdot)$ is the sigmoid function.

\textbf{SCD-DEC} As the forced alignment introduces additional computation cost, we can also adopt Transformer decoder structure to avoid explicitly aligning the acoustic features and text units, which predicts token-level speaker change probability $P_O = [p_0, \cdots, p_{U - 1}]$ directly conditioned on acoustic feature $Z_{\chi}$ and token embeddings $E_O = [\mathtt{Embed}(o_0), \cdots, \mathtt{Embed}(o_U)]$ from ASR. We write them as
\vspace{-0.5em}
\begin{equation}
\begin{aligned}
\bar{Z}_\chi & = \mathtt{TransformerEnc} \left(Z_{\chi} \right), \\
P_O & = \sigma \left(\mathtt{TransformerDec} \left(\bar{Z}_\chi, E_O \right) \right).
\end{aligned}
\vspace{-0.4em}
\end{equation}
Here we add several Transformer encoder layers to further process acoustic features $Z_{\chi}$ as $\bar{Z}_\chi$ instead of feeding them directly into Transformer decoder layers. As $P_O$ is a token-level probability vector, we map it to the final word-level results $P_W$ by choosing the maximum value of the token probability within each word.


\vspace{-0.7em}
\subsection{Objective Function}
\vspace{-0.3em}

Based on the previous discussions, the word-level (or token-level) speaker change detection is a typical classification problem with two classes. Considering the well-known sample balance issue, we adopt focal loss for the network optimization \cite{lin2017focal}. Given $Y_* = {\{Y_W, Y_O}\}$ as the speaker change labels on the word or token sequences, the loss function is wrote as
\vspace{-0.1em}
\begin{equation}
\mathcal{L} (P_*, Y_*) = -\sum_i \psi_i y_i \log (p_i) + \phi_i (1 - y_i) \log (1 - p_i),
\vspace{-0.7em}
\end{equation}
where $\psi_i = \alpha(1 - p_i)^\gamma$ and $\phi_i = (1 - \alpha) p_i^\gamma$. $\alpha$ and $\gamma$ are tuneable hyper-parameters to adjust the weight on positive \& negative examples and easy \& difficult samples, respectively.

\vspace{-0.5em}
\subsection{Chunk-wise Inference}
\vspace{-0.3em}

To address the SCD for long-form audio, we propose a chunk-wise processing scheme for our models. 
During runtime, each input long-form audio is firstly recognized by the streaming TT-ASR. 
Similar to the continuous speech separation in \cite{chen2020continuous}, the decoded word sequence is then segmented into overlapped chunks with $N_c$ words as chunk length, $N_h$ words as the look back and $N_f$ words as the future context. 
Later, speaker change points are estimated for each chunk by applying the proposed SCD networks that takes the word chunk and its corresponding acoustic feature frames as input. 
In SCD-DEC, the word emission time is employed to locate the boundary of acoustic feature for each chunk which is available once the beam search finishes. In SCD-ENC, the segmentation is applied directly on feature $Z_\chi^W$.
Finally, the chunk-wise decisions are merged to form the decision for the entire long-form audio.
In this way, the latency of the SCD network can be flexibly adjusted by changing the value of $N_h$, $N_c$ and $N_h$. Meanwhile, the ASR still remains a low latency setup regardless of the SCD setting. This flexibility allows the latency of SCD and ASR to be separately adjusted, which is beneficial to the performance of both tasks.

\vspace{-1em}
\section{Experiments Setup}
\vspace{-0.5em}

\subsection{Data}
\vspace{-0.3em}

The Microsoft in house data  was used to train the ASR and SCD models. The data included 65 thousand (K) hours of  anonymized speech with personal identifiable information removed. 
On-the-fly data simulation was performed to create the speaker turns for training of the SCD models. $M$ utterances from different speakers were concatenated together where $M$ is uniformly sampled from 2 to 4 for our proposed SCD models and 1 to 4 for joint SCD-ASR baselines.
The training labels $\{Y_W, Y_O\}$ in equation 7 are defined by assigning the last word and token position of each speaker utterance as 1 and other positions 0, respectively. 


An evaluation set containing 1.8 million (M) words, covering multiple application scenarios was constructed for ASR evaluation. 
For SCD, we collected two test sets: non-overlapped sessions from LibriCSS recording style (LS-ST) data \cite{chen2020continuous,wang2021continuous} and Microsoft vendor collected call center (MS-CC) data for evaluation. The LS-ST consisted of 114 16kHz recordings with a total duration of 13.6 hours. The MS-CC set included 400 8kHz audio sessions and the total duration was 24 hours. The total number of the speaker turns in LS-ST and MS-CC set were 4,834 and 9,409, respectively. During testing, the 8kHz audios were upsampled to 16kHz beforehand.

\vspace{-0.7em}
\subsection{Model Structures}
\vspace{-0.3em}

Our TT-ASR model used in this paper had a 2-layer convolutional network with kernel size 3 and stride size 2, a 18-layer Transformer encoder, a 2-layer 1024 dimensional LSTM predictor and a 512-dim joint network. The attention dimension of the MHSA layer in each encoder block was set to 512 with 8 heads and the 2048-dim FFN layer was adopted with the Gaussian error linear unit (GELU). 4035 word pieces are used as the ASR units together with blank and end-of-sentence symbol $\langle \text{eos}\rangle$. 

The joint SCD-ASR method in \cite{xia2022turn}  was selected as baseline in this paper where a speaker change symbol $\langle \text{sc}\rangle$ was added to the original TT-ASR output units to indicate the speaker change. Our proposed SCD-ENC model consisted of 6 non-causal Transformer encoders whose MHSA and FFN parameters were the same as TT-ASR's while SCD-DEC had a 2-layer non-causal Transformer encoder followed by 4 Transformer decoder layers. The total parameter size was 19.3M and 24.0M, respectively.


\vspace{-0.7em}
\subsection{Training and Evaluation Details}
\vspace{-0.3em}

The 80-dim log mel-filterbank using 25 msec window and 10 msec hop size was extracted as the input feature for TT-ASR.
Global mean and variance normalization as well as volume perturbation were applied on each utterance.
TT-ASR was trained on 16 GPUs for 300K steps with a regular linear decay learning rate scheduler. 25K warm-up steps was used and the peak learning rate was set to $1.6e^{-3}$.
We used chunk size 4 for the streaming mask \cite{chen2021developing} in TT-ASR, resulting a latency of 160 msec.
The context-independent (CI) phone branch \cite{zhao2021addressing} for forced alignment process was trained based on the 9-th encoder layer output ($Z_9$). 
In SCD-ASR baseline, speaker change token $\langle \text{sc}\rangle$ was inserted between the original transcriptions once there were multiple utterances sampled. 
For proposed SCD models, we performed training on 8 GPUs with 175K steps and 40K-frames batch size while freezing the ASR parameters. The peak learning rate was used as $1e^{-4}$ with 25K warm-up steps.  $\alpha$ and $\gamma$ in Equation 7 were set to 0.8 and 0.5, respectively. All the models in this work were optimized with the AdamW optimizer.
For streaming evaluation of proposed SCD network, we kept $N_h + N_c + N_f = 16$ in this work to illustrate the effectiveness of our method.


\vspace{-0.7em}
\subsection{Evaluation Metric}
\vspace{-0.3em}

F1 score was adopted to evaluate the quality of speaker change models and we used a similar way as \cite{chang2022turn}.
The speaker change denoted hypothesis and reference transcriptions were firstly formed, where a $\langle \text{sc}\rangle$ tag was inserted to the hypothesis and reference word sequence at every speaker change point. Then, the denoted reference and hypothesis were aligned and the F1 score was calculated
on those positions where the reference word was $\langle \text{sc}\rangle$.
For TT-ASR and SCD-ASR baselines, we directly used the output $\langle \text{eos} \rangle$ and $\langle \text{sc} \rangle$ symbol as speaker change decision for each sample, respectively. Regarding to SCD-ENC and SCD-DEC, the word position with a speaker change probability that is greater than a threshold of 0.5 was selected as the changing point.

\vspace{-1em}
\section{Results}
\vspace{-0.5em}

\subsection{Balance between SCD and ASR}
\vspace{-0.3em}

The SCD and ASR results of the proposed model and the baselines were shown in Table 1, where various observations can be made.
Although similar ASR performances were obtained by all systems, their performances in SCD tasks varied drastically. As TT-ASR was trained with only single speaker utterances, a  $\langle \text{eos} \rangle$ based speaker change detection failed to capture the characteristics of speaker difference, resulting in low F1 score for both data sets. 
With speaker turns included in the training data, a better F1 score was obtained by SCD-ASR while keeping a similar WER as TT-ASR. 
However, the F1 score is still far from satisfactory. One potential reason could be attributed to the low latency setting in streaming ASR, i.e., it is challenging to obtain high-quality speaker change detection under such low latency. Better F1 scores can be obtained for SCD-ASR by updating the parameters of the top 10 encoder layers. However, under this setup, the WER increased from 8.64\% to 8.75\%. This observation verifies that it is challenging for a tight integration between ASR and SCD to obtain optimum performances on both tasks.

On the other hand, both proposed methods achieved a significantly better SCD results than TT-ASR and SCD-ASR baselines. Specifically, the SCD-ENC achieved 91.3\% and 81.7\% in F1 score, which is 22.6\% and 14.3\% relative higher than the best SCD-ASR system, showing the efficacy of the proposed system. Within the proposed framework, the SCD-ENC demonstrated a slightly better performance over the SCD-DEC as the alignment based word boundaries were provided. However, SCD-DEC enjoyed a simpler implementation and a lower inference cost because the forced alignment step in SCD-ENC was usually computational costly. More discussions between two implementations are in the following sections.

\begin{table}[!t]
\setlength{\tabcolsep}{4pt}
\footnotesize
\centering
\caption{WER (\%) of our 1.8M test set and F1 score (\%) of LS-ST and MS-CC data set with TT-ASR and SCD-ASR. The best performance of our proposed SCD-ENC and SCD-DEC methods are also listed for the comparison.}
\begin{tabular}{c|c|c|c|c|c}
\toprule
\hline
\multirow{2}{*}{Method} & \multirow{2}{*}{Seed} & \multirow{2}{*}{Update ASR Enc.} & WER (\%) & \multicolumn{2}{c}{F1 (\%) $\uparrow$} \\ 
\cline{4-6}
 &  &  & 1.8M & LS-ST & MS-CC \\ 
\hline
TT-ASR & $\times$ & $\checkmark$ & 8.65 & 28.9 & 20.0 \\ \hline
\multirow{3}{*}{ SCD-ASR} & \multirow{3}{*}{ TT-ASR} & $\times$ & 8.80 & 54.2 & 38.8\\
 &  & $\checkmark$ & \textbf{8.64} & 60.7 & 66.7\\
 & & Top-10 & 8.75 & 74.5 & 71.5 \\
\hline
SCD-ENC & TT-ASR & $\times$ & 8.65 & \textbf{91.3} & \textbf{81.7} \\
\hline
SCD-DEC & TT-ASR & $\times$ & 8.65 & 86.4 & \textbf{81.7} \\
\hline
\bottomrule
\end{tabular}
\vspace{-6mm}
\end{table}

\vspace{-0.7em}
\subsection{Feature Investigation}
\vspace{-0.3em}

\begin{table}[!t]
\setlength{\tabcolsep}{5pt}
\footnotesize
\centering
\caption{F1 score (\%) on LS-ST and MS-CC data sets with our proposed methods. Different acoustic features and word boundary cues are explored. $N_c, N_h, N_f$ are set as $8, 4, 4$, respectively}
\begin{tabular}{c|c|c|c|c|c}
\toprule
\hline
\multirow{2}{*}{Method} & \multirow{2}{*}{Word Boundary} & \multicolumn{2}{c|}{Feature}  & \multicolumn{2}{c}{F1 (\%) $\uparrow$}  \\
\cline{3-4}
\cline{5-6}
 & & Acoustic & Text & LS-ST & MS-CC \\
\hline
\multirow{9}{*}{SCD-ENC} & \multirow{7}{*}{alignment} & $Z_0$ & $\times$  & 84.7 & 66.0 \\
& & $Z_4$ & $\times$ & \textbf{91.3} & \textbf{81.7}  \\
& & $Z_9$ & $\times$ & 88.2 & 76.9 \\
& & $Z_L$ & $\times$ & 83.8 & 71.3 \\
\cline{3-6}
& & $Z_*$ & $\times$ &  85.6 & 74.2 \\
& & $Z_\omega$ & $\times$ &  \textbf{90.5} & \textbf{80.3} \\
\cline{3-6}
& & $Z_4$ & $G_W$ & 91.2 & 80.4 \\
\cline{2-6}
& \multirow{2}{*}{emission} & $Z_4$ & $\times$ & 72.2 & 75.4 \\
& & $Z_4$ & $G_W$ & 76.0 & 76.4 \\
\hline
\multirow{7}{*}{SCD-DEC} &  \multirow{6}{*}{emission} & $Z_0$ & $E_O$  & 71.8 &  75.3 \\
 & & $Z_4$ & $E_O$ & 79.7 & \textbf{81.7} \\
 & & $Z_9$ & $E_O$ & 84.2 & 79.6 \\
 & & $Z_L$ & $E_O$ & 82.8 & 73.0 \\
 \cline{3-6}
 & & $Z_*$ & $E_O$ & 84.6 & 78.6  \\
& & $Z_\omega$ & $E_O$ & \textbf{86.4} & \textbf{81.7}  \\
\cline{2-6}
& alignment & $Z_\omega$ & $E_O$  & \textbf{90.4} & \textbf{82.3} \\
\hline
\bottomrule
\end{tabular}
\vspace{-6mm}
\end{table}

We compared the effectiveness of the acoustic feature candidates $Z_\chi \in \{Z_0, Z_4, Z_9, Z_L, Z_\omega, Z_*\}$ and the alternative semantic feature $G_W$ for SCD-ENC method, where the results were shown in Table 2. $Z_*$ means that we use fixed weights $\omega_i = \sum_l Z_l / L$.

In SCD-ENC, the shallow layer output $Z_4$ outperformed the other candidates, which verified that the shallow layer could capture more speaker characteristics. From $Z_4$ to $Z_L$, we observed the trend that the upper layer output gave worse F1 score, from 91.3\% and 81.7\% to 83.8\% and 71.3\%, which means that the best acoustic representation for ASR is not necessarily the best feature for SCD task. The weighted-sum feature $Z_\omega$ also showed competitive but slightly worse results than $Z_4$ and by analyzing the learnt weight of each layer, the shallow layers, especially the first four layers were assigned higher weights. Using fixed weights degraded the F1 scores significantly, indicating that the features of the different layers were not equally informative for the SCD task. Incorporating the text embedding features $G_W$ to the $Z_4$ based acoustic features didn't yield additional improvement. The pure acoustic features $Z_4$ were shown good enough for SCD-ENC network once the high-quality alignment based word boundaries were provided.

Regarding to SCD-DEC, $Z_\omega$ showed better performance than others on LS-ST data set and achieved the same F1 score on MS-CC data as $Z_4$. The weights of $Z_\omega$ had a similar distribution with the values learned by SCD-ENC, illustrating the shallow layer outputs contributed more in our word based SCD method. Feature $Z_0$ and $Z_L$ still gave worse results, degrading from 86.4\% \& 81.7\% to 71.8\% \& 75.3\% and 82.8\% \& 73.0\% , respectively. Compared with SCD-ENC, the best SCD-DEC model showed 5.4\% relatively worse F1 score on LS-ST data set. It was possibly due to the inaccurate audio chunking caused by the word emission time and we did analysis on this topic in the next section.


\vspace{-0.7em}
\subsection{Word Boundary Comparison}
\vspace{-0.3em}

In SCD-ENC, we adopted the accurate word boundaries from forced alignment to form the audio embedding feature for each word thus the performance highly relied on the quality of the provided boundary information. In Table 2, replacing the alignment based word boundary with the word emission time from beam search in SCD-ENC leaded to serious performance degradation, especially on LS-ST data set, from 91.2\% to 72.2\%, because of the latency between the real word ending time and the word emission time \cite{yu2021fastemit}. In this case, adding word embedding feature $G_W$ improved the F1 score to some extent but the effect was still limited. On the contrary, to figure out the reason that caused the performance gap between SCD-ENC and SCD-DEC, we tried to use the high-quality alignment based word timing for the audio chunking in SCD-DEC and get further improvement from 86.4\% and 81.7\% to 90.4\% and 82.3\% on LS-ST and MS-CC data sets, respectively. Thus if we can further reduce the latency of the word emission time, hopefully we can obtain similar performance as the SCD-ENC through SCD-DEC.

\begin{table}[!t]
\footnotesize
\centering
\caption{F1 score (\%) on LS-ST and MS-CC data sets with different history and future context size. We choose the best models in Table 2 (row 2 and 15) for the evaluation and constraint $N_h + N_f + N_c = 16$.}
\begin{tabular}{c|c|c|c|c}
\toprule
\hline
\multirow{2}{*}{$N_h$ \& $N_f$} & \multicolumn{2}{c|}{LS-ST}  & \multicolumn{2}{c}{MS-CC}  \\
\cline{2-5}
 & SCD-ENC & SCD-DEC & SCD-ENC & SCD-DEC \\
 \hline
0 & 80.0 & 79.5 & 73.6 & 76.6 \\
1 & 88.6 & 83.1 & 78.4 & 79.9 \\
2 & 89.8 & 84.4 & 80.0 & 80.8 \\
3 & 90.5 & 85.4 & 81.1 & 81.5 \\
4 & \textbf{91.3} & \textbf{86.5} & \textbf{81.7} & \textbf{81.7} \\
\hline
\bottomrule
\end{tabular}
\vspace{-6mm}
\end{table}

\vspace{-0.7em}
\subsection{Context Size for Inference}
\vspace{-0.3em}

The effect of the chunk size for inference was explored for both implementations and the results were shown in Table 3, where we kept $N_h + N_f + N_c = 16$ and chose $0\sim4$ for $N_h$ and $N_f$. Obviously compared with zero context results in the first row, even using small context, e.g., $N_h = N_f = 1$, can significantly improve the performance. And the trend was that the larger context we have, the better F1 scores were achieved but meanwhile, the higher computation cost we need to afford because of the larger overlap between chunks. We will try to reduce the model parameters and inference computation cost in the future work.



\vspace{-0.7em}
\section{Conclusions}
\vspace{-0.5em}

This paper presents a speaker change detection method with two implementations and they can work together with our previously studied streaming TT-ASR model asynchronously without losing ASR accuracy. The hidden output representations from the ASR encoder are shown to be the effective acoustic feature for SCD task. Compared with the prior joint SCD-ASR model, our method brings significant F1 score improvement on two data sets.
Our future work will explore the potential system fusion, the compression of the model size, the reduction of the computation cost and the extension of the multi-talker scenarios to support the speaker diarization task.

\vfill
\pagebreak
\bibliographystyle{IEEEbib}
\bibliography{main}

\end{document}